\documentclass[prb,aps,showpacs,epsfig]{revtex4}
\usepackage{epsf}

\def\E{{\rm e}}
\def\I{{\rm i}}
\def\D{{\rm d}}

\begin{document}

\title{Quantum  Dynamics of a Particle with a Spin-dependent Velocity}

\author{Claude Aslangul\footnote
  {{\bf e-mail:} aslangul@gps.jussieu.fr}}
\affiliation{Groupe de Physique des Solides, Laboratoire associ\'{e} au
CNRS (UMR 7588), \\ Universit\'e Paris 
Paris 6, Campus Boucicaut, 140 rue de Lourmel , 75015 Paris, France}


%
\begin{abstract}

We study the dynamics of a particle in continuous time and space, the
displacement of which is governed by an internal degree of freedom
(spin).  In one definite limit, the so-called quantum random walk is
recovered but, although quite simple, the model possesses a rich
variety of dynamics and goes far beyond this problem.  Generally
speaking, our framework can describe the motion of an electron in a
magnetic sea near the Fermi level when linearisation of the
dispersion law is possible, coupled to a transverse magnetic field. 
Quite unexpected behaviours are obtained.  In particular, we find that
when the initial wave packet is fully localized in space, the $J_{z}$ 
angular momentum component
is frozen; this is an interesting example of an observable which,
although it is not a constant of motion, has a constant expectation value. 
For a non-completely 
localized wave packet, the effect still occurs although less 
pronounced, and the spin keeps for ever memory of 
its initial state.  Generally speaking, as time goes on, the spatial density profile 
looks rather complex, as a consequence of the competition between drift 
and precession, and displays various shapes according to the ratio 
between the Larmor period and the characteristic time of flight. The 
density profile gradually changes from a multimodal quickly moving 
distribution when the scatttering rate is small, to a unimodal 
standing but flattening distribution in the opposite case.

\end{abstract}


\maketitle


\section{Introduction}\label{intro}

The term {\it quantum random walk} has been
coined some times ago by Aharonov and Davidovich \cite{Aharonov} to
qualify the motion of a quantum particle moving either left or right
according to the value of the $J_{z}$ component of its spin.  As such,
this is completely different of the quantum Brownian motion
problem, in which one fully describes in a quantum framework the
motion of a single (light) particle moving in a bath of heavy particles
-- a quantum version of the classical Brownian motion.  This latter
problem has been the subject of many papers in the past, revivified in
the 80's by the pioneering work of Caldeira and Leggett (for a review,
see  Leggett {\it etal} \cite{Caldelegggett}), showing that for the so-called ohmic
coupling with the bath, dynamical symmetry breaking occurs above a 
given threshold.  On a more elementary
level, simple models show how the classical brownian motion can be
generalized to the quantum world, although long time scales (and
algebraic behaviours) appear in correlation functions at small enough temperature, as 
shown by Aslangul {\it etal} 
\cite{Asl85}.  We shall adopt here the recent meaning as proposed by
Aharonov and Davidovich, shortened as QRW in the following.

This latter problem is presently the subject of an intense activity,
following the basic papers by Ambainis {\it et al} \cite{Ambainis01}, Nayak 
and Vishwanath \cite{nayak} and
D\"{u}r {\it et al} \cite{Duretal} (for a recent comprehensive review, see
the preprint by Kempe \cite{kempe081}).  The basic motivation lies in
the fact that QRW yields a dynamics at the opposite of classical
random walk: for a spin $\frac{1}{2}$ particle, and a symmetric
initial spin state with definite phases (see below for details), the probability density in
space $P_{\rm QRW}(x,\,t)$ displays two well-defined peaks, moving
away linearly in time from the starting place.  This is to be
contrasted with the $t^{1/2}$ spreading of the classical packet, the
particle staying (in the average) at its starting point when no drift
is present.  It can be said that QRW, except for oscillations in 
spatial density, displays the same basic features as the
classical non-difffusive motion with a drift (which can be obtained
from the {\it biased} diffusion equation $\frac{\partial P}{\partial
t}=- \vec\nabla [\vec v \,P-D\vec \nabla\,P]$ -- in the (singular)
limit $D\rightarrow\,0$, as explained for instance in the book by
Gardiner \cite{gardiner}) --, when the velocity itself is a random
variable assuming two values $\pm v_{0}$ with definite probabilities. 
This rather intuitive picture has been firmly grounded by the recent
work of Blanchard and Hongler \cite{blanchongler}.

Due to the quantum nature of the walk, all steps are strongly correlated (as contrasted 
to classical motion), as it is also the case for repeated mesurements 
yielding the Zeno paradox theoretically found by Misra and Sudarshan 
\cite{misrasudar} and observed by Itano {\it et al}\cite{itano}; this means 
that specific ever-lasting correlations are always 
relevant. One consequence is that the
linear dimension of the visited space increases linearly in time, {\it i. 
e.} much more rapidly for
QRW than for classical motion.  It is hoped, on a somewhat speculative
level, that Monte-Carlo algorithms could be improved by drawing
benefit of this fact (see {\it e. g.} Kempe \cite{kempe0205}). 

We adopt here a more general point of view. Indeed, we define a 
simple model containing a single parameter, noted $\alpha$ in the 
following, which is essentially the product of the spin-flip rate by 
the time of flight in space. As shown below, the limit $\alpha\rightarrow\,0$ can be viewed as 
the continuous space-time version of the conventional QRW. Out this 
limiting case, the model can describe the dynamics of an electron near 
the Fermi level (once the dispersion law has been linearized, 
$\varepsilon(k)\rightarrow\,\pm\hbar (k-k_{\rm F})v_{\rm F}$ as is 
done {\it e. g.} in the Luttinger model 
\cite{luttinger}), moving in a 
magnetic sea created by impurities, or coupled to nuclear spins 
(Vagner \cite{vagner}). Elastic collisions with the magnetic 
background 
can flip the spin of the moving electron and, simultaneously, change 
its velocity from $+v_{\rm F}$ to $-v_{\rm F}$. Alternatively, such an 
reversal can be induced by a transverse magnetic field forcing the spin to precess 
harmonically at the Larmor frequency $\omega$. 
As a whole, spin and orbital degrees of freedom are coupled, yielding 
intrication of the state as time goes on, and competition between 
precession and translation in space. This produces interesting 
effects; the first one (and probably the most unexpected) is freezing 
of the precession when the initial packet is quite narrow -- an effect 
which could have interesting applications in spintronics, as well in 
two-dimensional systems (McGuire {\it et al} \cite{mcguire}), in carbon 
nanotubes (Yang {\it et al}  \cite{yang}), in quantum dots (Levitov and Rashba 
\cite{levitov}) and in semiconductors (Dyakonov \cite{dyakonov}). A second 
characteristic feature is the $\propto t$ spreading of the wavepacket which arises in any 
case, even when the precession is extremely rapid: because of the 
latter, the particle has no time to choose between right and left 
moves, but dispersion still occurs, and very quickly as compared to 
classical random motion.

Obviously, the 
physical relevance of our model is subjected to small deviations 
around $k_{\rm F}$. On the other side, it is hoped that the basic (and surprising) results 
found below keep some relevance even with such restrictions, and at 
least can take place on space and time scales to be precised in real 
life. It is worthy to note that, provided such a physical picture is meaningful, 
$\alpha$ becomes an easily tunable parameter by varying 
either the magnetic field, or the filling-ratio of the band, or the 
density of the 
magnetic diffusing centers.

This paper is organized as follows.  We first define the model
(section \ref{qualitdisc}), we explicitely show the relation between
our model and QRW and we give a qualitative discussion of limiting 
cases. We then briefly derive the equations giving the
time-evolution of the density for an arbitrary spin $J$ (section
\ref{formaldens}), on which the spin-locking due to space confinement
in the $\alpha\,\rightarrow\,0$ limit can be directly shown.  This
fact is confirmed by a detailed study of the average value $J_{z}$
(section \ref{dynamicsJz}).  Then, we focus on the
$J=\frac{1}{2}$ case (section \ref{timeevoldens}), and calculate the
density profile $P(x,\,t)$, which displays many various shapes, some 
unexpected when
the two time-scales related to the Larmor precession and the
displacement in space are of the same order of magnitude.  Eventually,
conclusions are drawn and hints for future work are given.  In the
Appendix, details are given on the asymptotic analytical work required
to get insight on the features of the probability density.


\section{Model and qualitative discusssion}\label{qualitdisc}
For a spin $\vec J$ particle, our model Hamiltonian reads:
\begin{equation}
    H\,=\,\omega J_{y}+\hbar^{-1}\vec v.\hat {\vec p}\,J_{z}
    \label{defhamilton3d}\enspace.
\end{equation}
In this expression, $\omega$ is the spin-flip rate due to scattering 
on impurities, or the Larmor frequency due to a 
transverse magnetic field along the $y$ direction. $\vec v$ is the 
(scalar) quantity defining the velocity-scale, $\hat {\vec p}$ is the momentum (${\hat 
p}=-\I\hbar\vec \nabla$ in the $q$-representation). Despite some similitude at first 
glance, the Hamiltonian (\ref{defhamilton3d}) has nothing to do with 
the Dirac Hamiltonian, in which the momentum is coupled to the Dirac 
$\vec \alpha$ matrices; indeed, the angular momentum (spin) is there 
not given by $\vec \alpha$ and, in addition, the problem has here the 
dimensionality  $2J+1$ in spin space, instead of four in the Dirac 
theory. Also note that, in its one-dimensional form, with a space 
dependent $\omega$ and for $J=\frac{1}{2}$, this Hamiltonian has been used in various 
contexts such as inhomogeneous  supraconductors (de Gennes 
\cite{degennes}) and 
solitons in polyacetylene (Takayama {\it et al} \cite{makietal}).

In the following, we restrict to one-dimensional space; calling $Ox$ 
the line on which 
the particle moves, the Hamiltonian (\ref{defhamilton3d}) simplifies 
to:
\begin{equation}
    H=\omega J_{y}+\hbar^{-1}v\,{\hat p}\,J_{z}\enspace,\qquad 
    {\hat p}\,=\,-\I\hbar\frac{\partial}{\partial x}
    \label{defhamilton1d}\enspace.
\end{equation}
Note that the label $x$ of direction of motion and the three directions 
defining the components $J_{x},\,J_{y}$ and $J_{z}$ generally have no relations 
between them. Also note that the ``kinetic'' term is invariant under 
time-reversal.

\subsection{Continuous limit of the Quantum Brownian Walk}
\label{contlimitBrmotion}
In order to perform the continuous limit of QRW, let us recall the 
basic formalism. In the original model \cite{Duretal}, the 
spin-$\frac{1}{2}$ particle hops on a lattice (lattice spacing $a$, $n\in 
{\bf Z}$) from one site to the two 
first-neighbors,   
either right or left according to the value $+$ or $-$ of 
the 
$J_{z}$ component. In obvious notations, the operator $S$ generating this 
spatial  
motion is
\begin{equation}
    S\,=\,|+\rangle\langle +|\otimes\sum_{n}|n+1\rangle\langle n|
    +|-\rangle\langle -|\otimes\sum_{n}|n-1\rangle\langle n|
    \label{defhamiQRW}\enspace.
\end{equation}
Once a jump is done, the spin is changed by the action of a Hadamard 
matrix $T$ acting on the $|\pm\rangle$ spin states, which we choose of 
the following form:
\begin{equation}
    T\,=\,\frac{1}{\sqrt{2}}\,\left[\begin{array}{cc}
    1 & -1 \\
    1 & 1
    \end{array}\right]
    \label{defhadamard}\enspace.
\end{equation}
Thus, one step of the motion is generated by the product $TS$, 
which reflects the basic sequential nature of the walk.  
As a whole, for integer times $t$, the state $|\Psi(t)\rangle$ obeys the following 
equation:
\begin{equation}
    |\Psi(t+1)\rangle\,=\,TS|\Psi(t)\rangle\,\equiv\,H_{\rm QRW}|\Psi(t)\rangle
    \label{defhamilton1dQRW}\enspace.
\end{equation}
Performing now a Fourier transformation in space 
\mbox{($|k\rangle=N^{-1/2}\sum_{n}\E^{\I kna}$)}, one obtains:
\begin{equation}
    H_{\rm 
    QRW}\,=\,\sum_{k}\E^{-\I\hbar^{-1}\frac{\pi}{2}J_{y}}\,\E^{-\I\hbar^{-1}kaJ_{z}}|k\rangle\langle k|
    \label{defhamiQRWFou}\enspace.
\end{equation}
Let us now formally replace $t+1$ by $t+\Delta t$, $\frac{\pi}{2}$ by $\omega\Delta t$ in 
the first exponential; taking the limit $a\rightarrow\,0$, $\Delta 
t\rightarrow\,0$, $\frac{a}{\Delta t}=\mbox{Const}\equiv\,v$, one 
has:
\begin{equation}
    H_{\rm QRW}\,\rightarrow\,{\bf 1}+\frac{\Delta
    t}{\I\hbar}\sum_{k}(\omega J_{y}+kvJ_{z})|k\rangle\langle
    k|\label{defhamiQRWFoucont}\enspace.
\end{equation}
Going back to direct space, and taking now the limit $\lim_{\Delta t\rightarrow\,0}\frac{1}{\Delta 
t}[|\Psi(x,\,t+\Delta t)\rangle-|\Psi(x,\,t)\rangle]$, 
(\ref{defhamilton1dQRW}) yields:
\begin{equation}
    \I\hbar\,\frac{\partial }{\partial t}|\Psi(x,\,t)\rangle\,=\,H|\Psi(x,\,t)\rangle
    \label{schrodinger}\enspace
\end{equation}
where $H$ is the Hamiltonian given in (\ref{defhamilton1d}). This 
Schr\"{o}dinger 
equation retains the two essential features of QRW: the direction of 
the motion is determined by the value of the $J_{z}$ component, and the 
latter is not a constant of motion due to the transverse magnetic 
field (external or intrinsic) to which the particle is 
coupled though the operator $J_{y}$. 

Yet, a basic 
difference exists between $H_{\rm QRW}$ and $H$, due to the fact that in the discrete version, 
the particle jumps, and {\it then} the spin is changed by $T$: as 
already mentionned, the rules of 
the game are essentially sequential, allowing to state that the spin 
changes slowly as compared to the time of flight. By contrast, 
with $H$ given in (\ref{defhamilton1d}), both motion and spin-flip occur simultaneously. This means 
that, within the framework defined by $H$, one expects to recover the quantum 
random motion only when the Larmor frequency $\omega$ is quite small 
as compared to the time scale of the displacement.  As it stands, the 
Hamiltonian $H$ defines a model for which 
ordinary QRW is just one limit.

One additional 
ingredient is required, namely the initial state, which will be chosen of the 
spin-space separate 
form:
\begin{equation}
    |\Psi(x,\,t=0)\rangle\,=\,\psi(x)\otimes\sum_{M=-J}^{+J}c_{M}|M\rangle
    \equiv\,\psi(x)\otimes|\chi\rangle
    \label{initialstate}\enspace.
\end{equation}
$|M\rangle$ is the eigenstate of $J_{z}$ with the eigenvalue  
$M\hbar$; for physical purpose, $\psi(x)$ is chosen as a localized even function with a width $\sigma$, assuming real values in order 
to avoid any built-in inessential drift; for explicit calculations, 
we retain the gaussian normalized form:
\begin{equation}    
    \psi(x)\,=\,(\sqrt{2\pi}\sigma)^{-1/2}\,\E^{-x^{2}/(4\sigma^{2})}
    \label{gaussienne}\enspace.
\end{equation}

Obviousy, the state becomes intricate as time goes on, and one 
generally has at time $t>0$:
\begin{equation}
    |\Psi(x,\,t)\rangle\,=\,\sum_{M=-J}^{+J}\psi_{M}(x,\,t)|M\rangle
    \label{statetpos}\enspace.
\end{equation}
The main goal is to find the probability density in space, given by:
\begin{equation}
    P(x,\,t)\,=\,\sum_{M=-J}^{+J}|\psi_{M}(x,\,t)|^{2}
    \label{spatdensity}\enspace.
\end{equation}

The model is now completely defined, and involves just one 
dimensionless parameter, called $\alpha$ in all the following:
\begin{equation}
    \alpha\,=\,\frac{\sigma\omega}{v}
    \label{defalpha}\enspace.
\end{equation}
When $\alpha$ is large, the spin undergoes many Larmor precessions 
during a relatively small displacement in space; on the contrary, 
$\alpha$ small means that the spin precesses quite slowly when moving 
in space.

\subsection{The limiting cases}\label{limits}
Let us now briefly describe the limiting cases. As explained above, 
the $\alpha\rightarrow\,0$ limit must reproduce QRW. More precisely, 
when the Larmor period becomes much larger that any other relevant 
time-scale, the limit of the spatial density is the $(2J+1)$-modal distribution:
\begin{equation}
    \lim_{\alpha\,\rightarrow\,0}¥P(x,\,t)\,=\,\sum_{M=-J}^{+J}|c_{M}|^{2}\,|\psi(x-Mvt)|^{2}
    \quad\forall\,t\label{spatdensityalpnul}\enspace
\end{equation}
(for $J=\frac{1}{2}$, the two-peak splitting of ordinary QRW is recovered). 
Such a density trivially gives $\langle x\rangle 
(t)=vt\sum_{M}\hbar^{-1}\langle J_{z}\rangle (0)$ and $\Delta 
x^{2}(t)=(vt)^{2}\hbar^{-2}\Delta J_{z}^{2}(0)$. The $\propto t^{2}$ increase of its mean-square 
deviation merely reflects the fact that each peak of the density 
moves away linearly in time due to the persisting multimodal 
character of the density profile, which is frankly different from 
{\it spreading} in the usual sense. Note that, in the limit $\alpha=0$, the 
relative phases of the $c_{M}$ play no role, a symmetry which is broken 
for any finite $\alpha$: in the general case, 
this phases are essential (see below) and, in particular, determine 
whether the density $P(x,\,t)$ is symmetric in space or not.

On the other side, when $\alpha$ goes to infinity, spin and space
degrees of freedom become decoupled. Thus the spin stays immobile, and  simply 
precesses within its initial wave packet, which 
remains as it stood at the beginning:
\begin{equation}
    \lim_{\alpha\,\rightarrow\,+\infty}P(x,\,t)\,=\,|\psi(x)|^{2}
    \quad\forall\,t
    \label{spatdensityalpinf}\enspace.
\end{equation}
It will be seen in the following that these two trivial limits are 
indeed singular, especially the limit $\alpha\rightarrow\,0$ 
(because the velocity $v$ is in factor of the highest derivative in 
$H$); this can be considered as a first symptom of the richness of the dynamics 
for any {\it finite} $\alpha$. Anyway, the above limiting behaviours of $P(x,\,t)$   
 are expected to hold approximately true in the general case for times $t\ll 
\frac{2\pi}{\omega}$ and $t\ll\frac{\sigma}{v}$, respectively and 
mimick the actual dynamics.


\section{Formal expression of the density}\label{formaldens}
The Schr\"{o}dinger equation (\ref{schrodinger}) is formally easily 
solved by going to the $p$-representation. It then reads: 
\begin{equation}
    \I\hbar\frac{\partial}{\partial t}|\Phi(p,\,t)\rangle\,=\,\left(\omega
    J_{y}+\hbar^{-1}v\,p\,J_{z}\right)|\Phi(p,\,t)\rangle
    \label{schr-p}\enspace,
\end{equation}
where $|\Phi(p,\,t)\rangle$ is the $p$-representation of the state at
time $t$ ($p$ is now a scalar).  The time-evolution operator 
$U(p,\,t)$ is such that:
\begin{equation}
    |\Phi(p,\,t)\rangle\,=\,U(p,\,t)|\Phi(p,\,0)\rangle
    \label{propaonJqcq}\enspace,
\end{equation}
where $|\Phi(p,\,0)\rangle\equiv\phi(p)\otimes|\chi\rangle$ is the $p$-representation of the initial 
state (\ref{initialstate}); with the gaussian (\ref{gaussienne}), one 
has:
\begin{equation}
    \phi(p)=\left(\frac{2\sigma}{\hbar\sqrt{2\pi}}\right)^{\frac{1}{2}}\,
    \E^{-\sigma^{2}
 p^{2}/\hbar^{2}}
    \label{gaussrep}\enspace.
\end{equation}
By introducing the $p$-dependent unitary
transformation $R(p)=\E^{(\I\hbar)^{-1}\theta(p)J_{x}}$, $U(p,\,t)$ 
assumes the form:
\begin{equation}
    U(p,\,t)\,=\,R^{\dagger}(p)\,\E^{-\I 
    \hbar^{-1}\Omega(p)t\,J_{z}}\,R(p)
    \label{propagateurJqcq}\enspace,
\end{equation}
where:
\begin{equation}
    \Omega(p) 
    \,=\,[\omega^{2}+(v\hbar^{-1}p)^{2}]^{\frac{1}{2}}
    \label{defomega}\enspace,
\end{equation}
\begin{equation} 
    \cos\theta(p)\,=\,\frac{v\hbar^{-1}p}{\Omega(p)}\enspace,\quad
    \sin\theta(p)\,=\,\frac{\omega}{\Omega(p)}
    \label{defquprop}\enspace.
\end{equation}
This allows to write down the formal expression of the Fourier 
transform of the density probability, ${\cal 
P}(k,\,t)=\int_{-\infty}^{+\infty}\D x\,\E^{\I 
kx}P(x,\,t)$, as the following:
\begin{equation} 
    {\cal P}(k,\,t)=\int_{-\infty}^{+\infty}\D 
    p\,\,\phi^{*}_{+}\phi_{-}\langle\chi|U^{\dagger}(p+\frac{\hbar 
    k}{2},t)U(p-\frac{\hbar 
    k}{2},t)|\chi\rangle
    \label{foudens}\enspace,
\end{equation}
where $\phi_{\pm}=\phi(p\pm\hbar k/2)$. For 
arbitrary $J$, such 
an expression seems untractable as it stands, but it allows to look at 
the limiting case $\alpha\rightarrow 0$, which can be obtained by 
assuming a fully localized packet ($\sigma=0+$). Starting with the 
initial gaussian wavepacket (\ref{gaussienne}), a careful limiting procedure yields: 
\begin{equation} 
    \lim_{\alpha\rightarrow 0}{\cal P}(k,\,t)=\sqrt{\frac{2}{\pi}}
    \int_{-\infty}^{+\infty}\D 
    \xi\,\E^{-2\xi^{2}}\langle\chi|\E^{\I kvt\hbar^{-1}J_{z}}|\chi\rangle
    \label{foudensalphanul}\enspace.
\end{equation}
From this, one immediately obtains the limiting expression of the probability density in direct 
space:
\begin{equation} 
    \lim_{\alpha\rightarrow 
    0}P(x,\,t)=\sum_{M=-J}^{+J}\,|c_{M}|^{2}\,\delta(x-Mvt)
    \label{probdensalphanul}\enspace.
\end{equation}
This says that the initial narrow packet splits off in \mbox{$2J+1$} 
components, each of them going away from the starting point with its 
own velocity $Mv$. Expression (\ref{probdensalphanul}) holds true 
for any $J$ and any initial spin state.

This result is at first surprising (it would also trivially occur 
in the absence of magnetic background or of the transverse field ($\omega=0$) -- which also gives 
$\alpha=0$). It means that extreme confinement ($\sigma=0$) of the spin forces 
the component $J_{z}$ to have a constant expectation value, 
although $J_{z}$ is not a constant of motion due to the fact that 
$\omega\neq 0$. Obviously enough, one can question the validity of the 
above limiting procedure, because the value $\alpha=0$ is indeed singular; in 
fact, the above result, which provides an oversimplified picture of 
the spin-freezing phenomenon, can be easily confirmed by analysing an  
innocent-looking 
observable (indeed $J_{z}$ itself), directly obtained by the 
Heisenberg equations. This is done in the following section.



\section{Dynamics of the coordinate and of the component $J_{z}$}\label{dynamicsJz}
In order to get more insight in the above result, we now solve the 
Heisenberg equations of motion. They write:
\begin{equation} 
    {\dot x}_{\rm H}=v\hbar^{-1}J_{z\,{\rm H}}\enspace,\quad
    {\dot p}_{\rm H}=0
    \enspace,\quad{\dot J}_{z\,\rm H}=-\omega J_{x\,{\rm H}}
    \label{heisenbergeqmotxp}\enspace.
\end{equation}
\begin{equation} 
    {\dot J}_{x\,\rm H}=\omega J_{z\,{\rm H}}-\hbar^{-1}vp_{\rm 
    H}J_{y\,{\rm H}}\enspace,
    \quad
    {\dot J}_{y\,\rm H}=\hbar^{-1}vp_{\rm 
    H}J_{x\,{\rm H}}
    \enspace.
    \label{heisenbergeqmotJ}
\end{equation}
These equations can be readily integrated to yield:
\begin{equation} 
    J_{z\,\rm H}(t)=\,\vec e(t).\vec J
    \label{heisenintegreJ}\enspace,
\end{equation}
\begin{equation} 
    x_{\rm H}(t)=x+v\hbar^{-1}\vec T(t).\vec J
    \label{heisenintegrex}\enspace,
\end{equation}
where $\vec J$ (resp. $x$) coincides with $\vec J_{\rm H}(0)$ (resp. 
$x_{\rm H}(0)$ and where the components of the scalar vectors $\vec e(t)$ and $\vec 
T(t)$ are:
\begin{eqnarray} 
    e_{x}(t)=-\sin\theta\sin\Omega t\enspace,\hspace{50pt}\nonumber\\
    e_{y}(t)=\sin\theta\cos\theta(1-\cos\Omega t)\enspace,\hspace{15pt}\\    
    e_{z}(t)=\cos^{2}\theta+\sin^{2}\theta\cos\Omega t\enspace,\hspace{20pt}
    \label{heiseneuJ}
\end{eqnarray}
and $T_{u}(t)=\int_{0}^{t}e_{u}(t')\,\D t'$. 
The mean-square deviation of the coordinate is:
\begin{equation} 
    \Delta x^{2}(t)=\Delta x^{2}(0)
   +(v\hbar^{-1})^{2}\sum_{u,v} 
   (\langle T_{u}T_{v}J_{u}J_{v}\rangle-\langle 
   T_{u}J_{u}\rangle \langle T_{v}J_{v}\rangle)\enspace,
    \label{heiseneuDx2}
\end{equation}
where the brackets denote averages over the initial state 
(see (\ref{initialstate}) and (\ref{gaussrep})) -- remember that 
$\Omega$ and $\theta$ are functions of $p$, see (\ref{defomega}) and 
(\ref{defquprop}), and enter in a convolution with the Fourier transform 
of (\ref{gaussienne}). For any separate initial state, the averages 
factorize: $\langle T_{u}J_{v}\rangle=\langle T_{u}\rangle\langle J_{v}\rangle$ and so on.

These results allow a straightforward discussion displaying the 
strange features of the dynamics, especially the rather 
counterintuitive role of the initial spin 
state on the subsequent motion, especially on the symmetry of the 
probability density, as already discussed (see {\it e. g.} the 
analysis by Kempe \cite{kempe081}).

Close inspection first shows the essential role of the phases of the
coefficients $c_{M}$ appearing in the expansion (\ref{initialstate}).  Indeed, it is readily seen that when the
initial spin state is an eigenvector of $J_{y}$, then $\langle
J_{z}\rangle(t)$ and $\langle x\rangle(t)$ are constant (and thus
vanish at any time), whereas $\Delta x^{2}$ is still $\propto t^{2}$. 
Another consequence is that the density probability $P(x,\,t)$ is
symmetric in space in this case, and only in this case.  Thus, the dynamics,
and the parity of the spatial density, strongly depend on the relative
phases of the coefficients $c_{M}$ defining the initial spin state 
(see (\ref{initialstate})), not only of the weights
$|c_{M}|^{2}$ -- a feature which clearly separates the general 
$\alpha$-finite case
from the $\alpha=0+$ limit.  For any other preparation, $P(-x,\,t)\neq 
P(x,\,t)$ and the 
expectation values $\langle
J_{z}\rangle(t)$ and $\langle x\rangle(t)$ do vary in time, as 
examplified below for a definite preparation.

Indeed, let us assume that the initial spin state is the eigenstate 
$|M\rangle$ of $J_{z}$ ($J_{z}|M\rangle=M\hbar|M\rangle$); then, the expectation values at time $t$ 
are:
\begin{equation} 
    \langle J_{z}\rangle(t)=\,M\hbar\langle\cos^{2}\theta+\sin^{2}\theta\cos\Omega
    t\rangle\label{Jzmoyen}\enspace,
\end{equation}
\begin{equation} 
    \langle x\rangle(t)=\,Mv\,t\langle 
    \cos^{2}\theta+\sin^{2}\theta\frac{\sin\Omega t}{\Omega t}
    \rangle\label{coordmoyen}\enspace,
\end{equation}
\begin{equation} 
    \Delta x^{2}(t)=\sigma^{2}+
    v^{2}\left[\frac{1}{2}[J(J+1)-M^{2}]\langle T_{x}^{2}+T_{y}^{2}\rangle
    +M^{2}(\langle T_{z}^{2}\rangle-\langle T_{z}\rangle^{2})\right]
    \label{heiseneuDx2Jinit}
\end{equation}
In the preceding equations, $\langle\bullet\rangle$ denotes the 
average over the orbital variable: 
\begin{equation} 
    \langle\bullet\rangle=\,\sqrt{\frac{2}{\pi}}\frac{\sigma}{\hbar}
    \int_{-\infty}^{+\infty}(\bullet)\,\E^{-2(\sigma 
    p)^{2}/\hbar^{2}}\,\D p
    \label{moyennespp}\enspace.
\end{equation}

\vspace{0pt}\begin{figure}[htbp]
\centerline{\epsfxsize=240pt\epsfbox{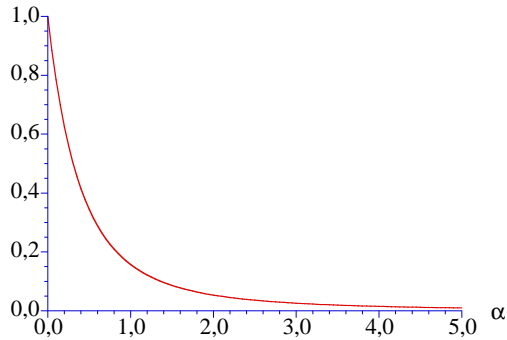}}
\vspace{0pt} \caption{Variation of the time 
average $\overline{\eta}$ (see (\ref{moytempJzexpl})) as a function of $\alpha$.}
\label{Jzmoyenalp}
\end{figure}

In particular, the expectation value of $J_{z}$ is given 
by \mbox{$\langle J_{z}\rangle(t)=\eta(t)\langle J_{z}\rangle(0)$} with:
\begin{equation} 
    \eta(t)=\,\sqrt{\frac{2}{\pi}}
     \int_{-\infty}^{+\infty}\, 
       \frac{\E^{-2\xi^{2}}}{\alpha^{2}+\xi^{2}}          
      \left(\xi^{2}+\alpha^{2}\cos\sqrt{\alpha^{2}+\xi^{2}}\,\frac{\omega 
      t}{\alpha}\right)\,\D\xi
    \label{moyJzexpl}\enspace
\end{equation}
(note that, with our definitions, $\frac{\omega 
t}{\alpha}=\frac{vt}{\sigma}$). $\eta(t)$ (obviously bounded by $\pm 
1$) is clearly  an oscillating function of time. This expression allows to discuss the unexpected behaviour of the 
average value of $J_{z}$ when $\alpha$ varies. First, let us look at 
the time-average value of $\eta$, $\overline{\eta}\equiv\lim_{t\rightarrow+\infty}\frac{1}{t}\int_{0}^{t}
    \eta(t')\D t'$. One readily finds:
\begin{equation} 
    \overline{\eta}=1-\sqrt{2\pi}\,\alpha\,\E^{2\alpha^{2}}[1-\Phi(\sqrt{2}\alpha)]
     \enspace
    \label{moytempJzexpl}
\end{equation}
where $\Phi$ is the probability integral \cite{gradryz}. 
$\overline{\eta}$ has the following behaviours:
\begin{equation}	
	\overline{\eta}\,\simeq\,\left\{ \begin{array}{ll}
	1-\sqrt{2\pi}\,\alpha & \mbox{\quad$\alpha\ll 1$} \\
	\frac{1}{4\alpha^{2}} & \mbox{\quad$\alpha\gg 1$}
	\end{array}\right.
	\label{asympetamoyt}	
\end{equation}

\vspace{0pt}\begin{figure}[htbp]
\centerline{\epsfxsize=320pt\epsfbox{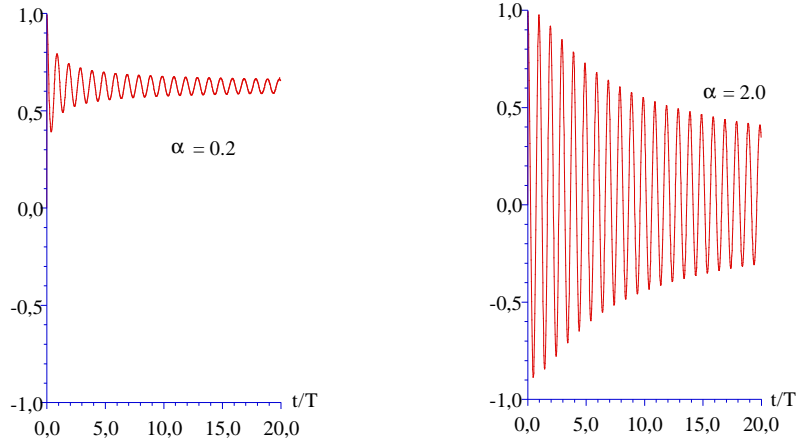}}
\vspace{0pt} \caption{Variation in time of $\eta(t)$ for a narrow 
initial wave packet (left) and for a large one (right). For an 
infinitely narrow wave packet, $\langle J_{z}\rangle$ remains constant. 
The unit of time is one period $T$ of the Larmor precession.}
\label{QuRdJzIOP}
\end{figure}

In addition, it is readily seen that $1-2\overline{\eta}\le\eta(t)\le 1$. 
Thus, for $\alpha$ small, $\eta(t)$ oscillates around a 
value which is quite close to one, showing that $\langle J_{z}\rangle$ 
becomes nearly constant in time. On the contrary, for $\alpha$ large, the oscillation 
takes place symmetrically around a quite small value. 
The variation of $\overline{\eta}$ as a function of $\alpha$ is 
given in Fig. \ref{Jzmoyenalp}.

At short times, one has $\eta(t)\simeq 1-\frac{1}{2}(\omega t)^{2}$. The behaviour of $\eta(t)$ at large times is easily found by using a 
saddle-point method. We find:
\begin{equation} 
    \eta(t)\simeq\overline{\eta}+\frac{2\alpha}{\sqrt{\omega 
    t}}\,\cos\left(\omega t+\frac{\pi}{4}\right)\enspace,\quad 
    t\gg \mbox{min }(\frac{\sigma}{v},\,\omega^{-1})
    \label{etaasympt}\enspace.
\end{equation}
The enveloppe of the oscillation around the time averaged value 
$\overline{\eta}$ thus 
decreases as $t^{-1/2}$. The variation of $\eta(t)$ at any time is plotted in Fig.\ref{QuRdJzIOP} for two 
values of $\alpha$.

These results confirm the confinement-locking of the spin: 
whereas $J_{z}$ is not a constant of motion, narrowing the width of 
the initial wave packet yields an expectation value which is less and 
less varying in time.

From (\ref{coordmoyen}), the average position of 
the particle is easily calculated and can be written as: 
\begin{equation} 
    \langle x\rangle(t)=Mvt\left[\overline{\eta}+\frac{\alpha^{3}}
    {\omega t}\sqrt{\frac{2}{\pi}}
    \int_{-\infty}^{+\infty}
    \frac{\E^{-2\xi^{2}}}
    {(\alpha^{2}+\xi^{2})^{3/2}}
    \sin\sqrt{\alpha^{2}+\xi^{2}}\,\frac{\omega t}{\alpha}\,\D \xi\right]
    \label{xmoyenMJz}\enspace.
\end{equation}
This represents a drift motion, with a damped-in-time ($\propto 
t^{-1}$) oscillation around the central position 
$Mvt\overline{\eta}$, imaging the forward/backward motion of the 
particle within its wavepacket as the spin precesses, causing inversion of the velocity.

\vspace{0pt}\begin{figure}[htbp]
\centerline{\epsfxsize=250pt\epsfbox{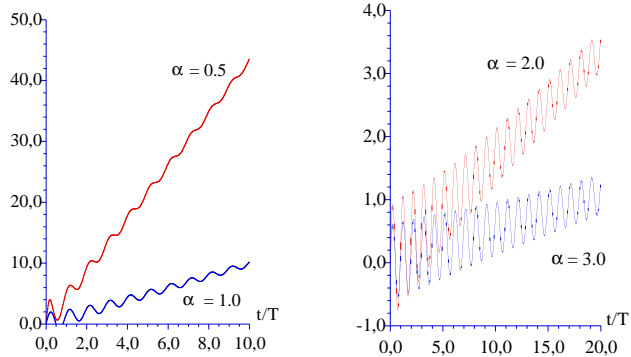}}
\vspace{0pt} \caption{Variation in time of $\langle x\rangle(t)$ for 
four values of $\alpha$, when the initial state is an eigenvector of 
$J_{z}$. Note the different horizontal and vertical scales.}
\label{QuRdxxIOP}
\end{figure}
It is readily seen that $\langle x\rangle(t)$ is bounded for 
any time and any $\alpha$; one finds:
\begin{equation} 
    Mvt[\overline{\eta}-(\omega t)^{-1}]<
    \langle x\rangle(t)<Mvt[\overline{\eta}+(\omega t)^{-1}]
    \label{xmoyenMJzborne}\enspace.
\end{equation}
This shows that, for such a preparation, the motion is always 
ballistic and simply approaches $\langle x\rangle(t)\simeq M\overline{\eta}vt$ at 
times which are large compared to the precession time scale. However, note that 
the effective velocity $\overline{\eta}v$ decreases rapidly when $\alpha$ 
becomes large (see (\ref{asympetamoyt}) and Fig. \ref{QuRdxxIOP}). In 
the asymptotic regime, one precisely has: 

\begin{equation} 
    \langle x\rangle(t)\simeq Mvt\left[\overline{\eta}+\frac{2\alpha}
    {(\omega t)^{3/2}}\sin \left(\omega t+\frac{\pi}{4}\right)\right]
    \label{xmoyenas}\enspace.
\end{equation}

Note that in  the limit $\alpha\rightarrow\,0$ but for an arbitrary initial spin 
state, one has 
$\langle J_{z}\rangle(t)=\,\langle J_{z}\rangle(0)$ and:
\begin{equation} 
\langle x\rangle(t)=vt\hbar^{-1}\langle J_{z}\rangle(0)
    \enspace,\qquad\Delta 
    x^{2}(t)=v^{2}t^{2}\hbar^{-2}\Delta J_{z}^{2}(0)
    \label{moyealpnul}\enspace,
\end{equation}
in agreement with (\ref{probdensalphanul}). The $\propto t^{2}$ 
increase of the mean-square dispersion merely reflects the fact that 
in this limit, the density is just the superposition of the $2J+1$ 
$\delta(x-Mvt)$ functions. 

Obviously, the mean square deviation gives a first feeling about the 
spatial density, but the qualitative discussion given above convinces 
one  
that a given $t^{2}$ increase of $\Delta x^{2}(t)$ can be realized in 
a variety of ways, ranging from two moving sharp peaks to a 
gaussian-like flattening {\it in situ}. Clearly,  a more precise analysis of 
the profile is required, and this is done in the following 
section for the $J=\frac{1}{2}$ case.


\section{Time evolution of the spatial density in the spin $\frac{1}{2}$ case}\label{timeevoldens}
In the $J=\frac{1}{2}$ case, the evolution operator can be easily 
written explicitely. After 
some algebra, we find the propagator $U(p,\,t)$ as the following:
\begin{equation}
    U(p,\,t)=\cos\frac{\Omega(p) t}{2}{\bf 1}-
    \I\sin\frac{\Omega(p) t}{2}\,\cos\theta(p)\,\sigma_{z}-\I\sin\frac{\Omega(p) 
    t}{2}\,\sin\theta(p)\,\sigma_{y}
    \label{propagateur}\enspace
\end{equation}
where the $\sigma_{u}$'s are the Pauli matrices. From this, one readily obtains the amplitudes $\psi_{\pm}(x,\,t)$, 
given by:
\begin{eqnarray}
    \psi_{\pm}(x,\,t)\,=\,\frac{1}{\sqrt{2\pi\hbar}}
    \int_{-\infty}^{+\infty}\D p\,\E^{\frac{\I}{\hbar}px}\nonumber\\
    \times\left[\left(\cos\frac{\Omega(p) t}{2}\mp\I\cos\theta(p)\sin\frac{\Omega(p) 
    t}{2}\right)\,c_{\pm}\mp \sin\theta(p)\sin\frac{\Omega(p) 
    t}{2}\,c_{\mp}\right]\,\phi(p)
    \label{amplitudes}
\end{eqnarray}
Close inspection of the  integrals in 
 (\ref{amplitudes}) again reveals that for any initial state which is 
 even and real, the density $P(x,\,t)=\sum_{\varepsilon=\pm}|\psi_{\varepsilon}(x,\,t)|^{2}$ is
 even only when \mbox{$|c_{+}|^{2}=|c_{-}|^{2}$} {\it and} 
 $\frac{c_{+}}{c_{-}}$ purely imaginary, {\it i.  e.} when the initial
 spin state is an eigenvector of $J_{y}$.  In all other cases,
 $P(-x,\,t)\neq P(x,\,t)$.

\vspace{0pt}\begin{figure}[htbp]
\centerline{\epsfxsize=500pt\epsfbox{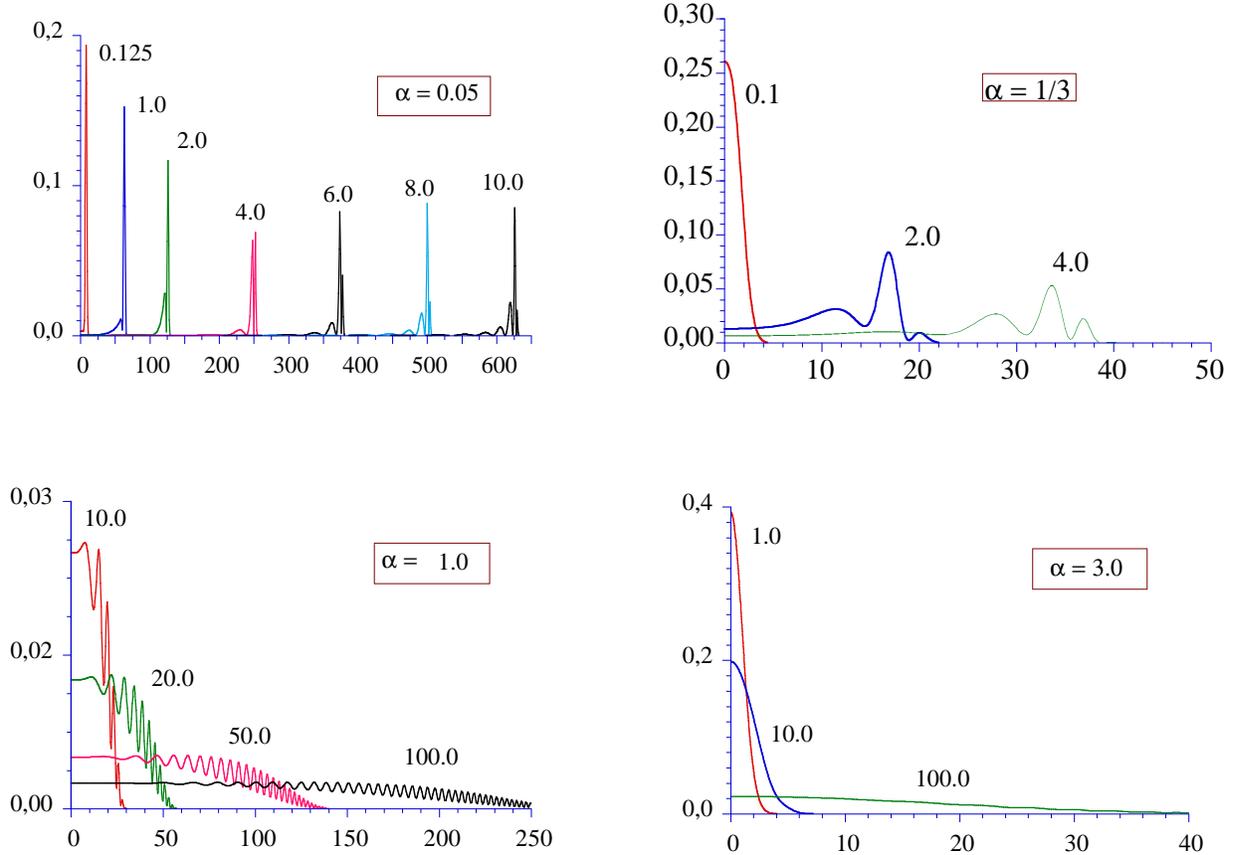 }}
\vspace{0pt} \caption{Snapshots of the (symmetric) density profile  at successive times $t$
for various values of the parameter $\alpha$; each curve is labelled by 
$\frac{t}{T}$ where $T$ is the Larmor period 
$\frac{2\pi}{\omega}$. The abscissa unit is the width $\sigma$ of the 
initial gaussian packet. Note the different scales from one drawing 
to another.}
\label{DensiteSymRecap}
\end{figure}


In the following, we restrict to the symmetric case, taking 
$c_{+}=\frac{1}{\sqrt{2}}$ and $c_{-}=\frac{\I}{\sqrt{2}}$, 
corresponding to $|\chi\rangle\,=\,|+\rangle_{y}$. Then,  the density can 
be written as follows:
\begin{equation}
    P(x,\,t)\,=\,\frac{1}{\sqrt{2\pi}\sigma}
    \sum_{r=1,\,2,\,3}|I_{r}(x,\,t)|^{2}
    \label{prob12}\enspace,
\end{equation}
where the three quantities $I_{r}$ are given integrals. With the 
gaussian wave packet (\ref{gaussrep}), the latter explicitely write ($X=\frac{x}{\sigma}$):

\begin{equation}
    I_{1}(x,\,t)\,=\,\frac{\alpha}{\sqrt{\pi}}
    \int_{-\infty}^{+\infty}\D \xi\,
    \frac{\E^{-\xi^{2}+\I\xi X}}{\sqrt{\alpha^{2}+\xi^{2}}}\,
    \sin\sqrt{\alpha^{2}+\xi^{2}}\frac{vt}{2\sigma}
    \label{int1}\enspace,
\end{equation}

\begin{equation}
    I_{2}(x,\,t)\,=\,\frac{1}{\sqrt{\pi}}
    \int_{-\infty}^{+\infty}\D \xi\,\E^{-\xi^{2}+\I\xi X}\,
    \frac{\xi}{\sqrt{\alpha^{2}+\xi^{2}}}\,
    \sin\sqrt{\alpha^{2}+\xi^{2}}\frac{vt}{2\sigma}
    \label{int2}\enspace,
\end{equation}

\begin{equation}
     I_{3}(x,\,t)\,=\, \frac{1}{\sqrt{\pi}}
    \int_{-\infty}^{+\infty}\D \xi\,\E^{-\xi^{2}+\I\xi X}\,
    \cos\sqrt{\alpha^{2}+\xi^{2}}\frac{vt}{2\sigma}    
      \label{int3}\enspace.
\end{equation}
These expressions allow an easy numerical calculation of the density 
$P(x,\,t)$ in various cases; the results are shown in Fig. \ref{DensiteSymRecap} 
and display the extreme variety of $P(x,\,t)$ when the single 
parameter $\alpha$ varies. For $\alpha$ small (remember this 
corresponds to QRW), the (two) peaks 
structure is clearly visible, and displays small satellites at the 
back of the packet. They are easily understood as arising from a 
(quantum) path in which the particle has undergone a small number of 
precessions. This is confirmed by the $\alpha=\frac{1}{3}$ curves: it 
can be checked that the number of peaks is close to 
$\frac{t}{T}$, where $T$ is the Larmor period. The extreme-right peak 
arises when the particle did not precess at all, then comes a secondary peak 
associated with one precession, and so on. When $\alpha$ increases, 
the structure is still present but diminishes quickly as time goes 
on. Eventually, for $\alpha$ large enough, the profile does not 
display any structure and vaguely looks like a standing gaussian wave 
packet. As a whole, the profile is rather sensitive to $\alpha$; also 
note that $P(x,\,t)$ remains notably non-zero  in the central region 
even at large times (see below, especially (\ref{Pcentrale})).

Approximate analytical expressions of the density 
can also be obtained (see Appendix). For $\alpha\ll 1$, we find that:
\begin{equation}
     P(x,\,t)\,\simeq\,\frac{\alpha^{2}}{16(2\pi)^{3/2}}
     \left|\int_{-\infty}^{+\infty}\,\D X'\,
     \E^{-\frac{1}{4}(X-X'-\frac{vt}{2})^{2}}H^{(2)}_{1}(\tilde X')
     \right|^{2}    
     \label{Pappttalpha}\enspace
\end{equation}  
where $\tilde X'=\alpha\sqrt{\frac{vt}{\sigma}|X'|}$ and where 
$H^{(2)}_{1}$ is the conventional Hankel function. Details given in 
the Appendix allow to understand that in this case ($\alpha\ll 1$), 
the two main peaks moving with the velocity $\frac{v}{2}$ are 
accompanied by small satellites, corresponding to those quantum paths 
with a few precessions.

On the other side, for $\alpha\gg 1$, one finds the plain gaussian 
distribution: 
\begin{equation}
     P(x,\,t)\,\simeq\,
     \frac{1}{\sqrt{2\pi} \Delta(t)}
     \E^{-\frac{x^{2}}{2\Delta(t)^{2}}}\enspace,
     \qquad\Delta\,=\,\sigma\,\left[1+\left(\frac{v^{2} 
     t}{4\sigma^{2}\omega}\right)^{2}\right]^{1/2}
     \label{Papgdalpha}\enspace.
\end{equation}
Because the precession is rapid, the particle does not move in the 
mean, but the effect is not averaged to zero and produces a linear in 
time increase of the width of the distribution.

Generally speaking, it appears that, for any 
$\alpha$, the central region remains populated due to the fact that 
$P(x=0,\,t)$ decreases rather slowly in time. Indeed, a stationnary-phase 
argument shows that:
\begin{equation}
     P(x=0,\,t)\,\simeq\,
     \frac{1}{\sqrt{2\pi} \Delta(t)}\,\sim 
     t^{-1}\quad\,\forall\,t\gg\mbox{Max 
     }\left(\frac{2\pi}{\omega},\,\frac{\sigma}{v}\right)
     \label{Pcentrale}\enspace.
\end{equation}

As for any initial state, the mean square dispersion of the coordinate always increases 
$\propto t$ at large times, and can be readily 
calculated for the symmetric case, using the results of section \ref{dynamicsJz}. 
In particular, one finds:  
\begin{equation}
     \lim_{t\rightarrow+\infty}\frac{\Delta x(t)}{t}
     \,=\,\frac{1}{2}\langle\cos^{4}\theta\rangle^{1/2}\,v 
     \,\equiv\,V(\alpha)\,v 
      \label{limdeltax}\enspace.
\end{equation}
with:
\begin{equation}
    \langle\cos^{4}\theta\rangle\,=\,1+\alpha^{2}-\sqrt{2\pi}\,
    \alpha(3+4\alpha^{2})\E^{2\alpha^{2}}\,[1-\Phi(\sqrt{2}\alpha)]    
      \label{vitdeltax}\enspace.
\end{equation}

\vspace{0pt}\begin{figure}[htbp]
\centerline{\epsfxsize=200pt\epsfbox{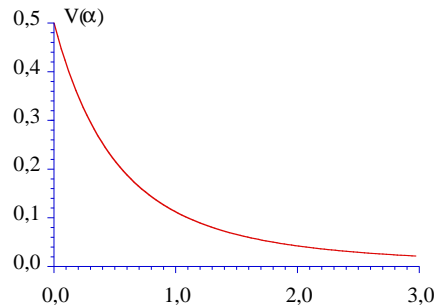}}
\vspace{0pt} \caption{Variation as a function of $\alpha$ of the 
velocity $V$ of the standard deviation $\Delta x(t)$ (see (\ref{asympvitdel})).}
\label{QuRdWkDeltaxmoy}
\end{figure}

\noindent $V(\alpha)$  is a monotonically decreasing function of $\alpha$ (see 
Fig. \ref{QuRdWkDeltaxmoy}), 
with the following behaviours:
\begin{equation}	
	V(\alpha)\,\simeq\,\left\{ \begin{array}{ll}
	\frac{1}{2}\left(1-3\sqrt{\frac{\pi}{2}}\,\alpha^{2}\right) & \mbox{\quad$\alpha\ll 1$} \\
	\frac{\sqrt{3}}{8\alpha^{2}} & \mbox{\quad$\alpha\gg 1$}
	\end{array}\right.
	\label{asympvitdel}	
\end{equation} 
This means that, apart from small damped in time oscillations, the 
motion is essentially ballistic, with a velocity $V(\alpha)$ going to zero 
when the precession frequency increases. With the reduced units 
used in Fig.\ref{DensiteSymRecap}, the width of the distribution is 
$2\pi/(\alpha V(\alpha))\,(t/T)$; rough estimates show that the gross  
features of the density profile are in agreement with the latter expression for the 
standard deviation.

\vspace{0pt}\begin{figure}[htbp]
\centerline{\epsfxsize=130pt\epsfbox{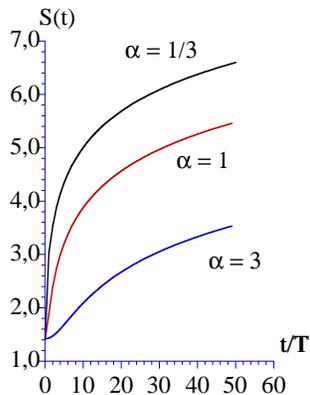}}
\vspace{0pt} \caption{Variation in time of the Shannon entropy $S(t)$ 
(left). The right part shows the ratio $S(t)/\ln t$, which becomes a 
constant at large times.}
\label{Entropie}
\end{figure}


As a final measure of the profile, let us analyse the Shannon 
entropy, defined as:
\begin{equation}	
	S(t)\,=\,-\int_{-\infty}^{+\infty}\D x\,P(x,\,t)\ln P(x,\,t)\enspace.
	\label{defentrShannon}	
\end{equation}
Three typical plots are given in the left part of Fig. 
\ref{Entropie}, showing that $S(t)$ -- an ever-increasing function in 
time --, changes gently with $\alpha$ and does not display a 
transition reflecting the bimodal/unimodal cross-over. In addition, it 
is seen that $S(t)\propto \ln\,t$ at large times, which means that, as 
it is often the case, $S(t)\propto \ln \Delta x(t)$.

\section{Conclusions}

In this paper, we presented a simple model describing the dynamics of
a particle with a linear dispersion law, when the direction of motion
is determined by the value of the spin, the former being able to flip
due to magnetic impurity scatttering or by coupling with an external
field.  The resulting intrication between orbital and spin degrees of
freedom yields a rather rich and complex dynamics with unexpected
features, governed by the single parameter $\alpha=\sigma\omega/v$ measuring the 
ratio between the time of flight and the Larmor period $T$.

The most surprising fact is spin-freezing when the width of the 
initial wave packet becomes quite small: narrowing the latter 
produces stronger and stronger  
shielding of the spin which can thus keep for ever the memory of its 
initial value. This robustness could be of interest in applications where 
the spin value carries sensitive information, {\it e. g.} in 
spintronics and in quantum spin computation.

Aside this remarkable fact, the density profile itself shows up a 
variety of shapes which reflect the competition between motion in 
space and Larmor precession. In the intermediate case where the 
characteristic time of flight $\sigma/v$ is of the order of the 
precession period, the profile displays a rich structure 
corresponding to the various quantum paths with zero, one, 
two,\ldots\, precessions: the number of rotations can be directly 
read by looking at the maxima of the density. In one extreme case 
($\alpha\ll 1$), one recovers conventional Quantum Random Walk in a 
space-time continuous framework, and its characteristic multimodal 
(bimodal for $J=1/2$) distribution. In the opposite case $\alpha\gg 1$, the 
particule hardly moves in the mean because of rapid precession, but 
the width of the packet increases proportionally to the time $t$.

Another interesting fact, already discussed in the past in the 
restricted QRW limit, is the role of the phases present in the initial 
spin state: it turns out that the space-density is even only when the 
initial state is an eigenvalue of the transverse spin $J_{y}$, {\it 
i. e.} when the coefficients have the same modulus {\it and} a 
definite phase relationship. For an infinitely narrow wave packet, 
this specific property disappears, and the subsequent motion becomes 
phase-independent: we do not have a simple explanation of this phase 
symmetry breaking.

In all cases, the RMS deviation of the coordinate $\Delta x(t)$ increases like $t$, much 
faster than in classical random motion. This universal increase 
integrates in fact various shapes, from sharp rapidly moving well 
defined peaks, to a standing flattening gaussian distribution. The 
same can be said about the Shannon entropy, which does not show up any 
cross-over when $\alpha$ varies; it essentially behaves like $\ln \Delta x$, as is often 
the case, and thus increases $\propto\,t$ at times large enough.

Some of our results are restricted to the $J=1/2$ case.  Obviously
enough, generalization to arbitrary $J$ is appealing; in particular,
it would be interesting to look at the high-$J$ (quasi-classical
limit), especially in order to analyse the above mentionned phase
symmetry-breaking phenomenon.  Furthermore, it would be interesting to
check whether the above shielding effect is robust against phase
decoherence, {\it i.  e.} to develop simple models incorporating coupling
to a quantum or classical bath.  Work in these directions is in
progress.

\section*{Appendix}

We now briefly sketch the methods allowing to obtain the approximate 
expressions (\ref{Pappttalpha}) and (\ref{Papgdalpha}) given in the 
main text. In all cases, the approximations start with the expression 
of the time-evolution operator $U$, and preserve unitarity.

Let us begin with the easiest case, namely $\alpha \gg 1$, {\it i. e.} 
when the spin precesses quickly and when the drift is slow. Then, the 
angle $\theta$ in  (\ref{defquprop}) is close to 
$\frac{\pi}{2}$. From (\ref{amplitudes}), one readily obtains:
\begin{equation}	
	\psi_{+}(x,\,t)\,\simeq\,\frac{1}{(2\pi)^{3/4}\sigma^{1/2}}
	\int_{-\infty}^{+\infty}\,\D k\,\E^{-k^{2}+\I k X}\,\E^{-\I\omega 
	t\sqrt{1+(k/\alpha)^{2}}/2}
	\label{psipgdalp}\enspace,	
\end{equation} 
and $\psi_{-}(x,\,t)=\I\,\psi_{+}(-x,\,t)$. Taking advantage of the gaussian 
cut-off, the square-root can be safely expanded when 
$\frac{k}{\alpha}\sim \frac{1}{\alpha}\ll 1$; the resulting gaussian 
integrals are readily evaluated, and one eventually finds the result 
given in (\ref{Papgdalpha}).

The other case, $\alpha\ll 1$, (swift drift ans slow precession) is
much more involved, due to the fact that the limit
$\alpha\rightarrow\,0$ is highly singular. From (\ref{defquprop}), the expression of 
$\psi_{+}$ is: 
\begin{equation}	
	\psi_{+}(x,\,t)\,=\,\frac{1}{(2\pi)^{3/4}\sigma^{1/2}}
	\int_{-\infty}^{+\infty}\,\D k\,\E^{-k^{2}+\I k X}\,
	\E^{-\I \frac{vt}{2\sigma}
	\,\sqrt{k^{2}+\alpha^{2}}}
	\label{psipetalp}\enspace,	
\end{equation}
and one still has $|\psi_{-}(x,\,t)|=|\psi_{+}(-x,\,t)|$. No simple 
approximation seems possible on such an expression but, since one 
expects that the rapidly moving peak aroud $x\sim \frac{vt}{2}$ be 
slighthy modulated by the precession, it is tempting to write an 
approximation using a convolution integal. For that purpose, we 
rewrite (\ref{psipetalp}) as follows:
\begin{equation}	
	\psi_{+}(x,\,t)\,=\,\frac{1}{(2\pi)^{3/4}\sigma^{1/2}}
	\int_{-\infty}^{+\infty}\,\D k\,\E^{\I k X}\,
	\E^{-k^{2}-\I k \frac{vt}{2\sigma}}\,
	\E^{\I \frac{vt}{2\sigma}
	\,(k-\sqrt{k^{2}+\alpha^{2}})}
	\label{psipetalp1}\enspace.	
\end{equation}
By the convolution theorem, this means that $\psi_{+}$ can be 
expressed as:
\begin{equation}	
	\psi_{+}(x,\,t)\,=\,\frac{1}{2^{3/4}\pi^{1/4}\sigma^{1/2}}
	\int_{-\infty}^{+\infty}\,\D X'\,
	\E^{-\frac{1}{4}\left(X-X'-\frac{vt}{2\sigma}\right)^{2}}
	F(X',\,t)
	\label{psipetalpconv}\enspace,	
\end{equation}
where $F(X,\,t)$ is the Fourier transform:
\begin{equation}	
	F(X,\,t)\,=\,\frac{1}{2\pi}
	\int_{-\infty}^{+\infty}\,\D k\,\E^{\I k X}\,
	\E^{\I \frac{vt}{2\sigma}
	\,(k-\sqrt{k^{2}+\alpha^{2}})}
	\label{defFFourier}\enspace.	
\end{equation}
\vspace{0pt}\begin{figure}[htbp]
\centerline{\epsfxsize=280pt\epsfbox{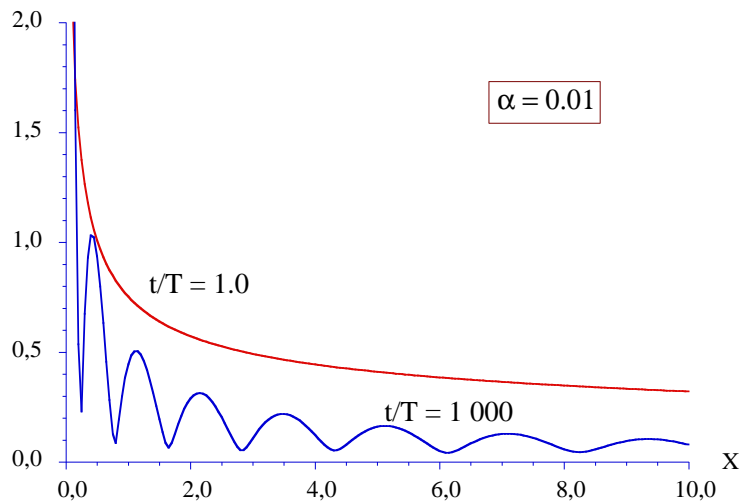}}
\vspace{0pt} \caption{Kernel $F(X,\,t)$ defined in (\ref{defFFourierfg})
for the convolution (see (\ref{psipetalpconv})).}
\label{MemConvol}
\end{figure}
Note that for $\alpha=0$, $F(X)$ reduces to the Dirac function 
$\delta(X)$. Now, we expand the (small) phase factor as 
$k-\sqrt{k^{2}+\alpha^{2}}\simeq  k-|k|(1+\frac{\alpha^{2}}{2k^{2}})$, 
allowing to write:
\begin{equation}	
	F(X,\,t)\,\simeq\,f(X,\,t)-f(X+\frac{vt}{2\sigma},\,t)
	\label{defFFourierfg}\enspace	
\end{equation}
where the function $f$ is defined as:
\begin{equation}	
	f(X,\,t)\,=\,\frac{1}{2\pi}\int_{0}^{+\infty}\,\D k\,
	\E^{\I X[k-\alpha^{2}vt/(4\sigma X k)]}
	\label{deffff}\enspace.	
\end{equation}
This function can be expressed with the Hankel functions 
$H^{(1,2)}_{\nu}(z)$
\cite{gradryz}. A somewhat tedious calculation yields:
\begin{equation}	
	f(X,\,t)\,=\,\frac{\alpha}{4}\mbox{sgn}(X)
	\sqrt{\frac{vt}{\sigma |X|}}
	\,H^{(2)}_{1}\left(\alpha\sqrt{\frac{vt}{\sigma}|X|}\right)
	\label{exprfff}\enspace.	
\end{equation}
By (\ref{defFFourierfg}), this completes the determination of the
kernel $F(X,\,t)$; nevertheless, analysis reveals that the second term 
in (\ref{defFFourierfg}) is 
quickly negligible, because of the rapid translation of each 
well-defined peak, so that one nearly always has $F(X,\,t)\,\simeq\,f(X,\,t)$.
Fig. \ref{MemConvol} shows the variation of 
$|F(X,\,t)|$ as a function of the reduced abscissa 
$X=\frac{x}{\sigma}$, at short and long times. In all cases, the 
strong peak in $F(X,\,t)$ near $X=0$ explains the persistence of the 
two 
main peaks in the density $P(x,\,t)$, whereas slowly decreasing 
($\sim X^{-1/2}$) oscillations are responsible for the secondary 
small peaks located just behind the main one. The minima arise from 
the zeroes of the $Y_{1}$ Bessel function included in $H^{(2)}_{1}$, 
which becomes denser and denser as time goes on; this can explain 
that at large times, the moving peaks for $\alpha\ll 1$ eventually 
grows up (see Fig. \ref{DensiteSymRecap}, upper left).


I am indebted to C. Caroli, R. Mosseri, P. Ribeiro and J. Vidal 
for helpful and numerous fruitful discussions.

\end{document}